\documentclass{article}
\usepackage{spconf,amsmath,graphicx}

\usepackage{amssymb}
\usepackage{mathtools}
\usepackage{algorithm}
\usepackage[noend]{algpseudocode}
\usepackage{multirow}
\usepackage{subfigure}
\usepackage{stackengine}
\usepackage{graphicx}

\title{BUILDING SYNTHETIC SPEAKER PROFILES IN TEXT-TO-SPEECH SYSTEMS}
\name{Jie Pu$^{1}$, Yixiong Meng$^{2}$, Oguz Elibol$^2$ 
}
\address{ $^1$Department of Engineering, University of Cambridge, UK\\
  $^2$Amazon Alexa, California, USA}
\begin{document}
%
\maketitle
\begin{abstract}
The diversity of speaker profiles in multi-speaker TTS systems is a crucial aspect of its performance, as it measures how many different speaker profiles TTS systems could possibly synthesize. However, this important aspect is often overlooked when building multi-speaker TTS systems and there is no established framework to evaluate this diversity. The reason behind is that most multi-speaker TTS systems are limited to generate speech signals with the same speaker profiles as its training data. They often use discrete speaker embedding vectors which have a one-to-one correspondence with individual speakers. This correspondence limits TTS systems and hinders their capability of generating unseen speaker profiles that did not appear during training. In this paper, we aim to build multi-speaker TTS systems that have a greater variety of speaker profiles and can generate new synthetic speaker profiles that are different from training data. To this end, we propose to use generative models with a triplet loss and a specific shuffle mechanism. In our experiments, the effectiveness and advantages of the proposed method have been demonstrated in terms of both the distinctiveness and intelligibility of synthesized speech signals. 
\end{abstract}
\section{Introduction}
\label{sec:intro}

With the advance of deep learning, modern text-to-speech (TTS) systems have developed end-to-end pipelines and enable the generation of speech signals approaching the human level of naturalness. For example, Tacotron-based approaches \cite{Wang2017TacotronTE} \cite{shen2018natural} will first map linguistic features of textual input into spectrograms, and then use a vocoder model \cite{oord2016wavenet} to obtain corresponding speech signals. Such an encoder-decoder network architecture and attention mechanism have been widely used now, and are shown to achieve remarkable performance in synthesized speech signals \cite{jia2018transfer}.

In this paper, we aim to solve a novel task in multi-speaker TTS systems: how to create new synthetic, fictional speaker profiles for the use of speech synthesis. The motivation behind this task is twofold: first and foremost, creating new synthetic speaker profiles helps to obtain a greater variety of voice profiles, which itself is a crucial aspect of multi-speaker TTS systems. Secondly, current multi-speaker TTS systems are limited to synthesize speech signals with the same speaker profiles as their training data. Such a limitation comes from the widely-used speaker embedding vectors \cite{skerry2018towards} \cite{gibiansky2017deep} in TTS systems. In particular, each speaker embedding vector captures the voice characteristics of one speaker in training data. This one-to-one correspondence between speaker embedding vectors and individual speakers in TTS systems posts constrains on its generalization capability. As a result, it struggles to generate speech signals with speaker profiles that are new and different from training data, thus hinders the diversity of voice profiles in multi-speaker TTS systems.

\begin{figure*}[tp]
  \centering
  {\includegraphics[width=0.95\hsize]{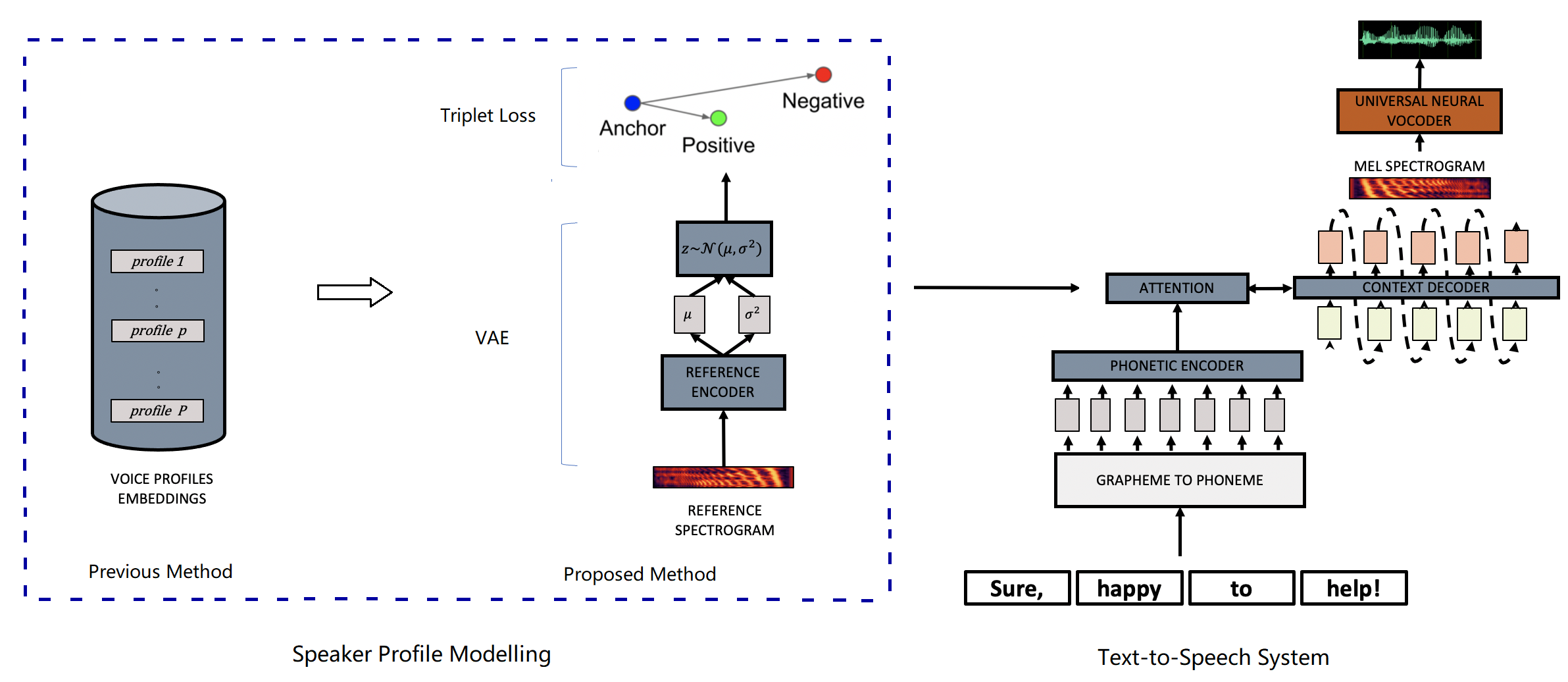}} 
  \caption{Overview of the proposed multi-speaker text-to-speech system. The proposed method focuses on the speaker profile modelling, which propagates the speaker profile information into generated speech signals. In previous methods, speaker profile information is represented as a fixed dimensional vector, called speaker/voice profile embeddings. In contrast, we propose to use a variational auto-encoder with the triplet loss to model speaker profiles. The obtained speaker profile is then conditioned on the attention layer of the context decoder, to generate speech signals of the corresponding speaker profile. } 
  \label{Overview}
\end{figure*}

To solve this problem, we propose to use generative networks for modelling speaker profiles in TTS systems, which breaks down the one-to-one correspondence between speaker embedding vectors and speaker identities. In particular, we choose to use a variational auto-encoder (VAE)  \cite{kingma2013auto} to synthesize speaker profiles along with TTS models. The trained generative model will learn the probabilistic distribution of speaker profiles and then create synthetic speaker profiles via sampling from this distribution. It is worth noting that the speaker profile modelling is decoupled from speech synthesis, in order to solely learn the probabilistic distribution of speaker characteristics. 

To enhance the quality of synthetic speaker profiles, we propose to use two specific techniques when training the VAE for speaker profiles: i) a triplet loss upon the latent space of VAE, in order to encourage a more structured latent space. Triplet loss \cite{chechik2010large} is a standard approach in metric learning and has shown promising performance in both face recognition \cite{schroff2015facenet} and speaker recognition \cite{chung2020in}, and ii) another challenge here is about disentanglement. During speech synthesis, many underline factors (e.g, speaker profiles, speech prosody and emotion status) are often mixed together. Disentangling these factors and explicitly generating certain speaker profiles is not easy. Therefore, we propose to use a shuffle mechanism of spectrograms to learn the disentangled representation of speaker profiles. During training, this mechanism will feed TTS models with spectrograms from the same speaker but have different speech content. The only consistency here is the speaker identity, and by doing so, speech characteristics other than speaker profiles will be filtered out.

\section{Methodology}

Our work is based on Amazon multi-speaker TTS system\footnote{https://aws.amazon.com/polly/}. In essence, we replace discrete speaker embedding vectors in the multi-speaker TTS system, and use a variational auto-encoder to learn speaker profile information from reference spectrograms. An overview of the proposed multi-speaker TTS system is depicted in Figure \ref{Overview}.

\subsection{Multi-speaker TTS system} 

Standard TTS systems nowadays use an end-to-end pipeline based on the sequence-to-sequence (seq2seq) modelling \cite{sutskever2014sequence} with an attention paradigm \cite{bahdanau2015neural}. It takes textual features as the input and produces spectrogram frames, which are then converted into waveforms. As shown in Figure \ref{Overview}, the TTS system includes an encoder, an attention-based decoder, and a neural vocoder \cite{shen2018natural}. 

To explicitly model speaker profiles in TTS, embedding vectors are used to represent each speaker in training data. These vectors will be concatenated into the attention layer of decoder, propagating the speaker profile information into synthesized speech signals. Specifically, these speaker embedding vectors can be initialized in different ways, such as using a uniform distribution over $[-0.1, 0.1]$ \cite{gibiansky2017deep}, the Glorot initialization \cite{glorot2010understanding} or extracted embeddings from speaker recognition and verification networks \cite{variani2014deep} \cite{heigold2016end} \cite{snyder2018x}. For those fixed-dimensional vectors that are extracted from neural networks, they are also widely used in the field of speaker recognition, with terminologies such as x-vector and d-vector.

\subsection{Proposed speaker profile modelling}

Herein we propose to use a variational auto-encoder to synthesize speaker profiles. As shown in Figure \ref{Overview}, an encoder of VAE is used as the reference encoder to obtain the speaker embedding vector $\textbf{z}$. It takes the input from a reference spectrogram and outputs the mean $\boldsymbol{\mu}$ and the variance $\boldsymbol{\sigma}$ of a 32-dimensional diagonal Gaussian distribution. As defined in \cite{kingma2013auto} the speaker embedding $\textbf{z}$: 

\begin{equation}
\label{eqn_z}
\begin{aligned}
& \textbf{z} = \boldsymbol{\mu} + \boldsymbol{\sigma} \odot \boldsymbol{\epsilon},\hspace{0.1cm} \text{where} \hspace{0.1cm} \boldsymbol{\epsilon} \sim \boldsymbol{\mathcal{N}}(\textbf{0}, \textbf{I})
\end{aligned}
\end{equation}
with $\odot$ we signify an element-wise product. $\boldsymbol{\epsilon}$ is sampled from a standard Gaussian distribution. After obtaining the speaker embedding $\textbf{z}$, it will be concatenated with the phonetic encoder output at the attention layer.

In the proposed method, synthetic speaker profiles are created by sampling from a 32-dimensional diagonal Gaussian distribution, where the latent variable $\textbf{z}$ lies in. With a well-trained VAE, the Gaussian distribution represents all speaker profile characteristics learned from training data. Sampling from it is equivalent to randomly choose one combination of these characteristics as the synthetic speaker profile. 

Besides, to enhance the quality of synthetic speaker profiles, we propose to use two specific techniques when training the VAE: i) a triplet loss and ii) a shuffle mechanism. 

\textbf{Triplet loss}. To encourage a structured latent space of speaker embeddings, we apply a triplet loss upon the latent variable $\textbf{z}$. The triplet loss will enforce embedding vectors of the same speaker become similar to each other, while embeddings of different speakers far away. As defined in \cite{chechik2010large}, the triplet loss can be formulated as follows: 
\begin{equation}
\label{eqn_triplet}
\begin{aligned}
& \resizebox{1\hsize}{!}{ $ \mathcal{L}(\textbf{A}, \textbf{P}, \textbf{N}) = max \left( \| f(\textbf{A}) - f(\textbf{P}) \|^2 - \| f(\textbf{A}) - f(\textbf{N}) \|^2  + \alpha, 0 \right) $} \\
\end{aligned}
\end{equation}
where \textbf{A} is an anchor input, i.e., a baseline utterance. \textbf{P} is a positive input, i.e., another utterance from the same speaker as \textbf{A}. \textbf{N} is a negative input, i.e., one utterance from a different speaker than \textbf{A}. $f$ is the VAE encoder that extracts embedding vectors from input utterances. The distance from the baseline (anchor) input to a positive input is minimized, while the distance from the baseline (anchor) input to a negative input is maximized. $\alpha$ is a margin between positive and negative pairs, where we use $\alpha = 0.5$ in the experiments.

\textbf{Shuffle mechanism}. One challenge here in modelling speaker profiles is how to disentangle the speaker profile information out of other mixed speech characteristics, such as speech prosody, environmental noise and emotional status. To this end, we apply a shuffle mechanism that dynamically selects the input reference spectrogram at each epoch. This mechanism will randomly choose the reference spectrogram from one of the output speaker's spectrograms. During training the only consistency between the reference spectrogram and the output spectrogram is their corresponding speaker identity. By doing so, speech characteristics other than speaker profiles will be filtered out after training. Compared to other approaches for disentangling representations in TTS \cite{hsu2018hierarchical} \cite{hsu2019disentangling}, the proposed method is both simple and effective as demonstrated in our experimental results.

\section{Experimental Evaluation}
\label{Experiments}

This section provides an experimental evaluation of the proposed method for building synthetic speaker profiles. First of all, it is worth mentioning that how to evaluate synthesized speech signals from TTS systems has always been a challenging research topic, and remains an open question nowadays. There are several subjective metrics used in past works to evaluate synthesized speech signals, such as the estimated mean-opinion scores (MOS), which relies on the judgement of human annotators. To promote the objective evaluation of synthesized speech signals, herein we propose to use three objective metrics summarized as follows:

\begin{itemize}
\item \textbf{Distinctiveness}. The proposed method is first assessed in the distinctiveness metric, which aims to quantify the diversity of synthetic speaker profiles and gives a global picture of its diversity. 
Specifically, we use the false-accept rate (FAR) obtained from a given speaker verification model as the distinctiveness: when the FAR is smaller, it indicates the speaker profiles of a set of synthesized speech signals are more diverse.

\item \textbf{Speaker similarity}. Different from the distinctiveness metric that provides a global picture of speaker diversity, we use cosine similarity scores between speaker profiles to conduct a local comparison. In particular, we examine the similarities of synthetic speaker profiles along an interpolation line, i.e., using interpolation to traverse between pairs of existing speaker profiles. It provides a zoom-in view on the learned VAE latent space, and check how the synthetic speaker profile is changing from one to the other.    

\item \textbf{Intelligibility}. Previous two metrics quantify how synthetic speaker profiles are different from each other, from both global and local perspectives. However, there exists a trivial solution to obtain distinctive speaker profiles: making synthetic speaker profiles as random noise (then they are indeed different from each other but is a meaningless solution). To prevent this, we use the intelligibility metric to measure the quality of  synthetic speaker profiles and synthesized speech signals. Similar to \cite{taylor2021confidence}, we use the word error rate (WER) of a given speech recognition model as the intelligibility: when the WER is smaller, it means the intelligibility of synthesized speech signals is better. 

\end{itemize}

The proposed method and its comparisons are trained on Amazon internal multi-speaker corpus, which contains more than 700k utterances from 2870 speakers. 8 V100 GPUs are used for training, and each model often takes $2 \sim 3$ days training time to converge. Besides, we use the ADAM optimizer to update model weights, with minimizing the $L_1$ loss between the original and generated mel-spectrograms.

\subsection{Distinctiveness} 
\label{Distinctiveness}
To quantify how synthesized speech signal is different from each other in term of its speaker profile characteristics, we use the false-accept rate (FAR) from the speaker verification model \cite{chung2020in} as the distinctiveness metric. It is worth emphasizing why FAR can be used here: let us assume the given speaker verification model is reasonably good, then a false acceptance case will happen (with an increase of FAR) when two compared speech signals have similar speaker profiles, i.e., their similarity score is high. In other words, it is hard for the speaker verification model to tell the difference between speech signals in term of their speaker identities. In that case, the FAR is large, which indicates the speaker profiles of speech signals are less distinctive. For a good performance, i.e., diversity of synthetic speaker profiles, we expect the value of FAR to be small.

Table \ref{FAR} reports the normalized FAR values of the proposed method and its comparison, calculated by dividing the FAR of a baseline method. As we can see, the proposed VAE outperforms d-vector that is used as speaker profile embeddings in multi-speaker TTS systems. To promote a more comprehensive view of FAR, there are three threshold percentile scores listed in the table. In all these thresholds, the proposed VAE with a triplet loss and a shuffle mechanism achieves the best performance, i.e, most diverse synthetic speaker profiles.

\begin{table}[!tb]
\centering
\caption{Normalized False Accept Rates (FAR) of the distinctiveness measurement. The FAR performance of d-vector is the baseline, and also the denominator for normalization. The best performance is in bold.}
\label{FAR}
\resizebox{\columnwidth}{!}{%
\begin{tabular}{|c|c|c|c|}
\hline
\multirow{2}{*}{Speaker profile modelling}           & \multicolumn{3}{c|}{Threshold percentiles} \\ \cline{2-4} 
                                 & 60th   & 70th   & 80th   \\ \hline
d-vector          & 1.000    & 1.000    & 1.000     \\ \cline{1-1}
Variational auto-encoder (VAE)                   & 0.802    & 0.831    & 0.945    \\ \cline{1-1}
VAE with a triplet loss & 0.734    & 0.686    & 0.836    \\ \cline{1-1}
VAE with a triplet loss and shuffle & \textbf{0.716}    & \textbf{0.657}    & \textbf{0.774}    \\
\hline
\end{tabular}%
}
\end{table}

\subsection{Speaker similarity} 

In this section, we aim to provide a zoom-in view on synthetic speaker profiles by comparing with existing speaker profiles in training data. Specifically, we take pairs of naturally spoken speech signals, encode them with the VAE encoder and then linearly interpolate on the latent space to obtain interpolated synthetic speaker profile vectors. The obtained speaker profile vector is concatenated into the attention layer of the TTS decoder to generate speech spectrograms of the corresponding speaker profile.

\begin{figure}[hb]
  \centering
  {\includegraphics[width=0.66\hsize]{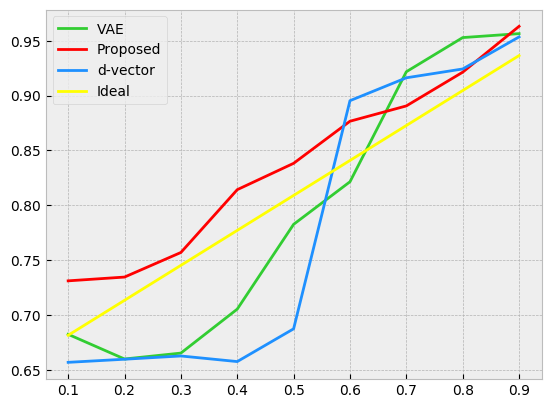}} 
  \caption{Cosine similarity scores between interpolated and existing speaker profiles. For a pair of existing speaker profiles, their encoded latent variables are $\textbf{z}_1$ and $\textbf{z}_2$. The interpolated speaker profiles are created by a weighted sum: $\textbf{z}_w = w*\textbf{z}_1 + (1-w)*\textbf{z}_2$, where the x-axis represents this weight $w$. For the y-axis, it shows the cosine similarity between $\textbf{z}_w$ and $\textbf{z}_1$. The label \textit{Proposed} represents the use of VAE with a triplet loss and shuffle, and the \textit{Ideal} stands for the most ideal case (with only theoretical possibility) of manifold smoothness.} 
  \label{Similarity}
\end{figure}

We use cosine similarity scores from the speaker verification network \cite{chung2020in} to measure the similarity/difference between  speaker profiles. As shown in Figure \ref{Similarity}, the interpolated synthetic speaker profiles from the proposed method show a smooth and gradual transformation from one natural speaker profile to the other, which demonstrates the continuity of the learned manifold space $\textbf{z}$. Compare to the baseline approach using d-vector (where you can see a sudden jump from the weight $0.5$ to $0.6$), our synthetic speech signals also have more diverse speaker profiles, which further supports the result about distinctiveness in section \ref{Distinctiveness}. 
 
\subsection{Intelligibility} 
\label{Intelligibility}
To evaluate the intelligibility of synthesized speech signals, we use the word error rate (WER) from Amazon speech recognition system\footnote{https://aws.amazon.com/transcribe/}. As shown in \cite{taylor2021confidence}, ASR-based metrics, e.g. WER, can perform reliably for evaluating intelligibility of TTS systems, and indeed on a comparable level to the paid human annotators/listeners.

Table \ref{WER} shows the normalized WER results of the proposed method and its comparison, calculated by dividing the WER of a baseline method. Herein, we compare different methods using a certain number of synthetic speaker profiles generated from these methods. As we can see, the proposed VAE with a triplet loss achieves the best performance with the smallest WERs, showing the benefit of the triplet loss during training, as it enforces a more structured latent space on $\textbf{z}$. On the other hand, the shuffle mechanism has a negligible effect on the intelligibility of synthesized speech signals.

\begin{table}[!tb]
\centering
\caption{Normalized Word Error Rates (WER) of the intelligibility measurement. The WER performance of d-vector is the baseline, and also the denominator for normalization. The best performance is in bold.}
\label{WER}
\resizebox{\columnwidth}{!}{%
\begin{tabular}{|c|c|c|c|}
\hline
\multirow{2}{*}{Speaker profile modelling}           & \multicolumn{3}{c|}{Number of speaker profiles} \\ \cline{2-4} 
                                 & 1                  & 250                & 500               \\ \hline
d-vector         & 1.000              & 1.000               & 1.000              \\ \cline{1-1}
Variational auto-encoder (VAE)                   & 0.926      & 0.983               & 0.924              \\ \cline{1-1}
VAE with a triplet loss & \textbf{0.914}      & 0.971      & \textbf{0.878}     \\ \cline{1-1}
VAE with a triplet loss and shuffle & \textbf{0.914}      & \textbf{0.970}      & 0.879     \\ \hline
\end{tabular}%
}
\end{table}

\section{Conclusion}
In this paper, we propose to use generative models to synthesize speaker profiles in TTS systems. Specifically, a variational auto-encoder is used to learn the probabilistic distribution of speaker profiles, with a triplet loss to regularize its latent space and a shuffle mechanism to disentangle speaker information. By doing so, the proposed method enables TTS systems to generate synthetic speaker profiles that have not been seen in training data. The effectiveness of the proposed method has been demonstrated in our experiments.

\newpage
\bibliographystyle{IEEEbib}
\bibliography{mybib}

\begin{thebibliography}{10}

\bibitem{Wang2017TacotronTE}
Yuxuan Wang, R.~Skerry-Ryan, Daisy Stanton, Y.~Wu, Ron~J. Weiss, Navdeep
  Jaitly, Z.~Yang, Y.~Xiao, Z.~Chen, S.~Bengio, Quoc~V. Le, Yannis
  Agiomyrgiannakis, R.~Clark, and R.~A. Saurous,
\newblock ``Tacotron: Towards end-to-end speech synthesis,''
\newblock in {\em INTERSPEECH}, 2017.

\bibitem{shen2018natural}
Jonathan Shen, Ruoming Pang, Ron~J Weiss, Mike Schuster, Navdeep Jaitly,
  Zongheng Yang, Zhifeng Chen, Yu~Zhang, Yuxuan Wang, Rj~Skerrv-Ryan, et~al.,
\newblock ``Natural tts synthesis by conditioning wavenet on mel spectrogram
  predictions,''
\newblock in {\em 2018 IEEE International Conference on Acoustics, Speech and
  Signal Processing (ICASSP)}. IEEE, 2018, pp. 4779--4783.

\bibitem{oord2016wavenet}
Aaron van~den Oord, Sander Dieleman, Heiga Zen, Karen Simonyan, Oriol Vinyals,
  Alex Graves, Nal Kalchbrenner, Andrew Senior, and Koray Kavukcuoglu,
\newblock ``Wavenet: A generative model for raw audio,''
\newblock {\em arXiv preprint arXiv:1609.03499}, 2016.

\bibitem{jia2018transfer}
Ye~Jia, Yu~Zhang, Ron Weiss, Quan Wang, Jonathan Shen, Fei Ren, Patrick Nguyen,
  Ruoming Pang, Ignacio~Lopez Moreno, Yonghui Wu, et~al.,
\newblock ``Transfer learning from speaker verification to multispeaker
  text-to-speech synthesis,''
\newblock in {\em Advances in neural information processing systems}, 2018, pp.
  4480--4490.

\bibitem{skerry2018towards}
RJ~Skerry-Ryan, Eric Battenberg, Ying Xiao, Yuxuan Wang, Daisy Stanton, Joel
  Shor, Ron Weiss, Rob Clark, and Rif~A Saurous,
\newblock ``Towards end-to-end prosody transfer for expressive speech synthesis
  with tacotron,''
\newblock in {\em International Conference on Machine Learning}, 2018, pp.
  4693--4702.

\bibitem{gibiansky2017deep}
Andrew Gibiansky, Sercan Arik, Gregory Diamos, John Miller, Kainan Peng, Wei
  Ping, Jonathan Raiman, and Yanqi Zhou,
\newblock ``Deep voice 2: Multi-speaker neural text-to-speech,''
\newblock in {\em Advances in neural information processing systems}, 2017, pp.
  2962--2970.

\bibitem{kingma2013auto}
Diederik~P. Kingma and Max Welling,
\newblock ``Auto-encoding variational bayes,''
\newblock in {\em 2nd International Conference on Learning Representations,
  Canada, April}, 2014.

\bibitem{chechik2010large}
Gal Chechik, Varun Sharma, Uri Shalit, and Samy Bengio,
\newblock ``Large scale online learning of image similarity through ranking.,''
\newblock {\em Journal of Machine Learning Research}, vol. 11, no. 3, 2010.

\bibitem{schroff2015facenet}
Florian Schroff, Dmitry Kalenichenko, and James Philbin,
\newblock ``Facenet: A unified embedding for face recognition and clustering,''
\newblock in {\em Proceedings of the IEEE conference on computer vision and
  pattern recognition}, 2015, pp. 815--823.

\bibitem{chung2020in}
Joon~Son Chung, Jaesung Huh, Seongkyu Mun, Minjae Lee, Hee~Soo Heo, Soyeon
  Choe, Chiheon Ham, Sunghwan Jung, Bong-Jin Lee, and Icksang Han,
\newblock ``In defence of metric learning for speaker recognition,''
\newblock in {\em Interspeech}, 2020.

\bibitem{sutskever2014sequence}
Ilya Sutskever, Oriol Vinyals, and Quoc~V Le,
\newblock ``Sequence to sequence learning with neural networks,''
\newblock {\em Advances in neural information processing systems}, vol. 27, pp.
  3104--3112, 2014.

\bibitem{bahdanau2015neural}
Dzmitry Bahdanau, Kyunghyun Cho, and Yoshua Bengio,
\newblock ``Neural machine translation by jointly learning to align and
  translate,''
\newblock in {\em 3rd International Conference on Learning Representations,
  ICLR 2015}, 2015.

\bibitem{glorot2010understanding}
Xavier Glorot and Yoshua Bengio,
\newblock ``Understanding the difficulty of training deep feedforward neural
  networks,''
\newblock in {\em Proceedings of the international conference on artificial
  intelligence and statistics}, 2010, pp. 249--256.

\bibitem{variani2014deep}
Ehsan Variani, Xin Lei, Erik McDermott, Ignacio~Lopez Moreno, and Javier
  Gonzalez-Dominguez,
\newblock ``Deep neural networks for small footprint text-dependent speaker
  verification,''
\newblock in {\em 2014 IEEE International Conference on Acoustics, Speech and
  Signal Processing (ICASSP)}. IEEE, 2014, pp. 4052--4056.

\bibitem{heigold2016end}
Georg Heigold, Ignacio Moreno, Samy Bengio, and Noam Shazeer,
\newblock ``End-to-end text-dependent speaker verification,''
\newblock in {\em 2016 IEEE International Conference on Acoustics, Speech and
  Signal Processing (ICASSP)}. IEEE, 2016, pp. 5115--5119.

\bibitem{snyder2018x}
David Snyder, Daniel Garcia-Romero, Gregory Sell, Daniel Povey, and Sanjeev
  Khudanpur,
\newblock ``X-vectors: Robust dnn embeddings for speaker recognition,''
\newblock in {\em 2018 IEEE International Conference on Acoustics, Speech and
  Signal Processing (ICASSP)}. IEEE, 2018, pp. 5329--5333.

\bibitem{hsu2018hierarchical}
Wei-Ning Hsu, Yu~Zhang, Ron~J Weiss, Heiga Zen, Yonghui Wu, Yuxuan Wang, Yuan
  Cao, Ye~Jia, Zhifeng Chen, Jonathan Shen, et~al.,
\newblock ``Hierarchical generative modeling for controllable speech
  synthesis,''
\newblock in {\em International Conference on Learning Representations}, 2018.

\bibitem{hsu2019disentangling}
Wei-Ning Hsu, Yu~Zhang, Ron~J Weiss, Yu-An Chung, Yuxuan Wang, Yonghui Wu, and
  James Glass,
\newblock ``Disentangling correlated speaker and noise for speech synthesis via
  data augmentation and adversarial factorization,''
\newblock in {\em ICASSP 2019-2019 IEEE International Conference on Acoustics,
  Speech and Signal Processing (ICASSP)}. IEEE, 2019, pp. 5901--5905.

\bibitem{taylor2021confidence}
Jason Taylor and Korin Richmond,
\newblock ``Confidence intervals for asr-based tts evaluation,''
\newblock {\em Proc. Interspeech 2021}, pp. 2791--2795, 2021.

\end{thebibliography}
\end{document}